\documentclass[aps,amsfonts,twocolumn,unsortedaddress,nofootinbib]{revtex4}
\usepackage{epsfig}
\usepackage{graphicx}
\usepackage{amsmath}
\begin{document}
\title{A note on DSR-like approach to space-time}
\author{R.  Aloisio}
\affiliation{INFN - Laboratori Nazionali
del  Gran  Sasso,  SS.  17bis,  67010  Assergi  (L'Aquila)  -  Italy}
\author{A.  Galante}
\affiliation{INFN - Laboratori Nazionali
del  Gran  Sasso,  SS.  17bis,  67010  Assergi  (L'Aquila)  -  Italy}
\affiliation{Dipartimento di Fisica,  Universit\`a di L'Aquila, Via
Vetoio  67100 Coppito (L'Aquila) - Italy}
\author{A.  Grillo}
\affiliation{INFN - Laboratori Nazionali
del  Gran  Sasso,  SS.  17bis,  67010  Assergi  (L'Aquila)  -  Italy}
\author{E.  Luzio}
\affiliation{Dipartimento di Fisica,  Universit\`a di L'Aquila, Via
Vetoio  67100 Coppito (L'Aquila) - Italy}
\author{F.  M\'endez}
\affiliation{INFN - Laboratori Nazionali
del  Gran  Sasso,  SS.  17bis,  67010  Assergi  (L'Aquila)  -  Italy}

\begin{abstract}
In  this  note we  discuss  the  possibility  to define  a  space-time
with  a DSR  based approach.  
We show  that the strategy of defining a non linear realization 
of the Lorentz symmetry with a consistent vector composition
law cannot be reconciled with the extra request of an invariant 
length (time) scale.
The latter request forces to abandon the group structure of the
translations and leaves a space-time structure where points
with relative distances smaller or equal to the invariant scale can not
be unambiguously defined.
\end{abstract}

\maketitle
\section{introduction}
It is widely believed that  the space-time, where physical process and
measurements take place, might have a structure different from a
continuous and differentiable manifold, when it is probed at the Planck
length  $\ell_p$.   For example,  the  space-time  could  have a  foamy
structure  \cite{foam}, or  it  could be  non-commutative  in a  sense
inspired by  string theory  results \cite{noncom} or  in the  sense of
$\kappa$- Minkowski approach \cite{kamin}. 

If this happens in the space-time, in the momentum space there must also
be a scale, let say $p_p$, that signs this change of structure of the
space-time, even if the interplay  between length and momentum ( $p\sim
c \hbar \lambda^{-1}$) will presumably change when we approach 
such high energy scales. 

One could argue that, if the Planck length gives a limit at which one
expects that quantum gravity effects become relevant, then it would be
independent from  observers, and one  should look for  symmetries that
reflect this property.   Such argument gave rise to  the so called DSR
proposals,  that is,  a deformation  of the  Lorentz symmetry  (in the
momentum space) with two invariant  scales: the speed of light
$c$ and $p_p$ (or $E_p$) \cite{amelino,smolin,mag}. 

In this note,  we will discuss  this class  of deformations of the
Lorentz symmetry and its realization in the space-time.  Approaches to
the problem inspired by the momentum space formulation  have  been presented
\cite{mag,spacetime}, but  our approach is quite  different from these
because we demand the existence  of an invariant measurable
physical scale compatible with
the  deformation of  the composition  law of  space-time  coordinates 
induced by the non linear transformation.

It has  also been  claimed  that $\kappa$-Minkowski \cite{KM} 
gives  a  possible realization of the DSR   principles  
in space-time \cite{KMDSR},   however  the
construction is still not satisfactory since 
 the former is only compatible with momentum composition law 
non symmetric under the exchange of particles
 labels (see discussions in \cite {nsim}).
 In this work we are dealing with non
 linear realizations of the Lorentz algebra
 which induce symmetric composition law 
and therefore it is not
 compatible with the $\kappa$-Minkowski approach.
 
The main results of our studies are: {\it i})  the strategy of 
defining a non linear realization 
of the Lorentz symmetry with a consistent vector composition
law cannot be reconciled with the extra request of an invariant 
length (time) scale;
{\it ii}) the request of an invariant length
forces to abandon the group structure of the
translations and leaves a space-time structure where points
with relative distances smaller or equal to the invariant scale can not
be unambiguously defined.

In  the next  section we  will  explore the  approach to  DSR in  the
momentum  space and  will  implement these  ideas in  the space-time
sector. 
In the  final   section  conclusion  and  discussions  are
presented.

\section{Non   Linear  realization  of   Lorentz  group   approach  to
 space-time}
In this  section we will first review the  approach to  DSR as a  non linear
realization  of  the  Lorentz  transformations in  the  energy-momentum
space,  and then try to apply these  ideas to  the space-time.  For a
more general review  of DSR see for example  \cite{rev} and references
therein. 

DSR  principles are realized in the energy-momentum
space  by  means  of  a   non-linear  action  of  the  Lorentz  group
\cite{smolin,luk}.  More precisely, if the coordinates of the physical
space $ P$ are $p_\mu = \{ p_0,{\bf p}\}$, we can define a non- linear
function $F: P\rightarrow {\cal P}$,  where $\cal P$ is the space with
coordinates $\pi_\mu=  \{\epsilon, {\boldsymbol \pi}\}$, on  which the Lorentz
group acts  linearly. We  will refer  to $\cal P$  as a  
{\em classical} momentum space.

In terms of the previous variables,  a boost of a single particle with
momentum \footnote{From  here we will omit all  the indexes.
When necessary we will  use the convention  $a^\mu=(a,{\bf  a})$} 
$p$ to  another
reference frame, where  the momentum of the particle is $p'$, is given
by
\begin{equation}
\label{dsrboost}
p'= F^{-1}\circ\Lambda\circ F~ [p] \equiv
{ \cal B }[p].
\end{equation}

Finally, an addition  law ($\hat +$) for momenta,  which is covariant
under the action of $\cal B$, is
\begin{equation}
\label{dsrsum}
p_a\hat{+}p_b=F^{-1}\left[ F [p_a]+F[p_b]\right],
\end{equation}
and satisfies ${ \cal B }[p_a\hat{+}p_b] = {\cal B}[ p_a] \hat{+} {\cal
B}[ p_b]$ . 

In this formulation, the requirement of having an invariant scale 
fixes the action of $F$ on some points of the real space $P$.  Indeed,
since the Lorentz
transformation leaves the points  $\pi=0$ and $\pi=\infty$ invariant,
one  sees  that if  we  demand  invariance  of Planck  momentum  (${\bf
  p}_{p}$)  (or   energy  ($E_{p}$))  then   $F[{\bf  p}_{p}]$  (or
$F[E_{p}]$)  only can  be  $0$ or  $\infty$  \cite{smolin}.  A  general
discussion on the possible deformations is given in \cite{luki}. 

In the following we explore the possibility to extend
the above discussion generalizing it to define
a non linear realization of Lorentz symmetry in space-time.
In  analogy with the  momentum space,  we will  consider a  {\em real}
space-time  $X$ with  coordinates $x^\mu=\{x^0,\cdots,x^3\}$  and will
assume:  $a$) the existence  of  an auxiliary  space-time  $\cal X$  with
coordinates  $\xi^\mu=\{\xi^0,\cdots,\xi^3\}$   (called  {\em classical
space-time}),  where  the  Lorentz  group  acts linearly  and $b$)  the
existence of an invertible map $G[x]$ such that $G[x]:X \rightarrow {\cal X}$. 

Boosts in the space-time will be defined in the same way as DSR boosts
in the momentum space, that is
\begin{equation}
\label{boostx}
{\cal B}= G^{-1}\circ \Lambda \circ G,
\end{equation}
where $\Lambda$ is the Lorentz boost, which acts linearly on $\cal X$. 

As was done in the energy-momentum space, we want to define a space-time 
vector composition law covariant under the action of deformed boost:
\begin{equation}
\label{suma}
x_a \hat{+} x_b = G^{-1}[G[x_a]+G[x_b]].
\end{equation}
This definition implies that a vector can always be written as
the sum of two (or more) vectors and this decomposition is
covariant under boosts:
\begin{equation}
\label{sumas}
\hat{\delta}_{(ab)}=G^{-1}\left[G[x_a]-G[x_b] \right]
= \hat{\delta}_{(ac)}~\hat{+}~\hat{\delta}_{(cb)},
\end{equation}
for any $x_c$.
With relation (\ref{suma}) we can define the operation of translation of
all vectors of our space by a fixed vector $\alpha$:
\begin{eqnarray}
\label{trasla}
{\hat T}_\alpha(x)&\equiv&x\hat{+}\alpha.
\end{eqnarray}
With the above definition the translations behave as in the
standard case under the action of boosts:
\begin{eqnarray}
{\cal B}\left({\hat T}_\alpha(x)\right)
&=&
{\hat T}_{{\cal B}(\alpha)}\left({\cal B}(x)\right). 
\end{eqnarray}
It is easy to check that these transformations form  a group.
We will call $e$ the neutral element of the group $i.e.$ the
neutral element for the composition law. From the relations
(\ref{suma}),(\ref{trasla}) we read that the neutral element
is such that $G[e]=0$ $i.e$ its image corresponds to the origin
of the {\em classical} space.

\subsubsection{Invariant scales}
With  the  previous  definitions we want to add the condition
that, under the deformed boosts, an invariant measurable physical 
scale (both a time or length scale) has to exist.
In doing that we have in mind that, eventually, this scale
is related with the Planck length (or time).
Let us call $\hat{\delta}_{p}$ the vector which defines the
invariant scale. By this we mean that $\hat{\delta}_{p}$ is any
vector of the form
\begin{equation}
\label{timescale}
\left(
\begin{array}{cc}
T_{p}\\
\bf x
\end{array}\right)
\qquad\qquad
\mbox{or}
\qquad\qquad
\left(
\begin{array}{cc}
t\\
{\bf x}_p
\end{array}\right),
\end{equation}
where ${\bf x}_p$ is any spatial vector with modulus equal to the 
invariant scale and $T_p$ is the invariant time scale. 
The invariance condition we impose is that a time 
(length) equal to the invariant scale is not affected by boosts.
This corresponds to the physical intuition that
any Planck-length segment (whatever his time position) or Planck-time
interval (whatever his space position) should remain unaffected
by a change in reference frame:
$i.e.$
\begin{equation}
\label{planckboost}
{\cal B}
\left(
\begin{array}{cc}
T_{p}\\
\bf x
\end{array}\right)
=
\left(
\begin{array}{cc}
T_{p}\\
{\bf x}' 
\end{array}\right)
\quad
\mbox{and}
\quad
{\cal B}
\left(
\begin{array}{cc}
t\\
{\bf x}_p
\end{array}\right)
=
\left(
\begin{array}{cc}
t'\\
{\bf x}_p
\end{array}\right).
\end{equation}
In the {\em classical} space we have the invariant 
$(\xi^0)^2-({\boldsymbol \xi})^2$
whose image in the physical space is obviously invariant under the
action of deformed boost. This, together 
with the requirement of invariance under rotations, allows to
demonstrate that the above relations have to satisfy $t=t'$ and 
${\bf x}={\bf x}'$.

The above result is valid if we assume the existence of $i$) only a temporal
invariant scale, $ii$) only a spatial invariant scale, $iii$) both
a temporal and a spatial invariant scale.

For
example, if  we assume  an invariant time  scale $T_p$, the vector
$(T_p,{\bf x})$ (with arbitrary ${\bf x}$) under the
action of a boost transformation has to be modified in such a way to 
keep $T_p$ fixed as well as to leave the Casimir 
$C(T_p,{\bf x})=(\xi^0)^2-({\boldsymbol \xi})^2=(G^0(T_p,{\bf x}))^2-
({\bf G}(T_p,{\bf x}))^2$ invariant. 
Since, in physically interesting cases, $C(T_p,{\bf x})$
depends on the spatial coordinates only via the modulus $|{\bf x}|$, 
we get $C(T_p,|{\bf x}|)=C(T_p,|{\bf x'}|)$ $i.e.$
$|{\bf x'}|=|{\bf x}|$. We get the interesting result that,
assuming only a temporal invariant, all the vectors with time
component equal to the invariant quantity (and any ${\bf x}$) 
keep the modulus of the spatial coordinate unchanged when boosted. 
Clearly this does not imply that any length
scale is an invariant one since, in general, 
${\cal B}(t,{\bf x})=(t',{\bf x}')$ with ${\bf x} \ne {\bf x}'$ if
$t\ne T_p$.
The same analysis shows that, if we assume the existence of a length
invariant, the Casimir invariant implies 
${\cal B}(t,{\bf x}_p)=(t,{\bf x}_p)$ for any $t$ value.

 The final result is that our invariant
vector(s) has to satisfy the following relation
\begin{equation}
\label{invarianza}
{\cal B}\hat{\delta}_{p}=\hat{\delta}_{p},
\end{equation}
or, equivalently,
\begin{equation}
\label{critic}
\Lambda G[\hat{\delta}_{p}]=G[\hat{\delta}_{p}].
\end{equation}

The invariant points of standard Lorentz transformations 
are $0$ and $\infty$ 
($i.e.$ vectors where all the components are zero or infinity)
and we can only use one of this two points to ensure the condition
(\ref{invarianza}).
In DSR in the momentum space we have a similar situation:
in that case the fixed point at infinity is used to guarantee
invariance of (high) energy or momentum scales.
Since we demand our model to be equivalent to usual space-time
when the distances (times) are much bigger than the invariant
scale we expect $G$ to approach the identity in this conditions
and this implies $G[\infty]=\infty$.

Then the only possibility left is that the image  of the invariant 
vectors under $G$  must be $0$. This condition is equivalent to 
the condition we used to define the neutral element of translations
$e$.
This result indicates that at this point we are getting in
troubles: infinitely many different physical vectors are mapped
to the same vector in the {\em classical} space $i.e.$ the 
function $G$ can not be any more invertible when we approach
the invariant scale.
For example if we consider two distinct 
points $x_a$, $x_b$ and assume
that their distance (in space and/or time) equals the
invariant quantity we have to write
\begin{equation}
\label{iguali}
x_b \hat{-} x_a = \hat{\delta}_{p}
\end{equation}
and, after applying the $G$ function to previous relation, we get
$G[x_a]=G[x_b]$ in contrast with our assumption of $G$ invertible.
We conclude that, to satisfy the invariance condition,
we loose the uniqueness of the
neutral element of translations and therefore its group structure.

\begin{figure}[t!]
\begin{center}
\includegraphics[width=0.32\textwidth,angle=270]{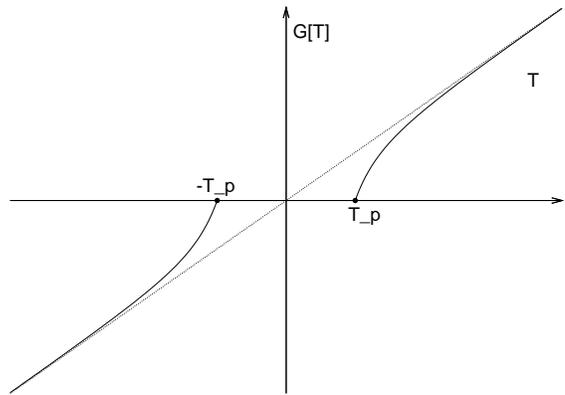}
\caption{An example of the function $G_0$}
\label{G}
\end{center}
\end{figure}
In figure 1 we give an explicit realization of the temporal
part of the $G$ function
for a vector with a fixed spatial part and a varying temporal 
component
in order to represent it as an ordinary one-dimensional function.
As required, we recover the standard 
Lorentz transformation for large times $i.e.$ the $G$ function
becomes the identity for $t>>T_{p}$ while goes to $0$ once  
the invariant scale is reached.
It is clear that both the images of $T_{p}$ and $-T_{p}$ are zero: 
we loose the uniqueness ($G$ is not invertible at this points)
and an invertible extension of $G$ can not be defined on points 
in the interval $[-T_p,T_p]$.

What is happening is that, to enforce a physical invariant, 
we had to choose a $G$ function such that all the
vectors of the {\em classical} space (the vertical axis in
fig. 1) map, in the physical space (the horizontal axis in fig. 1),
into vectors with spatial or temporal length larger than the 
physical invariant scale. When we try to consider vectors in
the physical space with spatial or temporal length equal or
smaller than the invariant scale we have no counterpart in
the {\em classical} space. 
For example, if we want to consider a vector $x$ that is half
of the invariant vector we should write
\begin{equation}
\label{meta}
x \hat{+} x = \hat{\delta}_{p}
\end{equation}
and this equation corresponds to $G[x]=-G[x]$ or, equivalently,
$G[x]=0$, $i.e.$ $x=\hat{\delta}_{p}$.

With a speculative attitude we can imagine that the model suggests
that a process of space (or time) measurement can never give a
result with a precision better than the invariant length since
all points that differ by an invariant length are practically
indistinguishable.

In the particular example of fig. 1, if $T$ is a time 
interval measured in a given space point, we have to conclude
that we can not speak of $\|T\|\le T_p$ or, equivalently,
this  can   be  understood as  an
indetermination  in the  measurement  of  time when  the  Planck scale  
is reached. Therefore we can imagine this signals an intrinsic
obstruction  to measure  a time (distance)
when it becomes of the order of the invariant scale.

Two comments are in order here. We have demonstrated the
impossibility of implementing a  minimal scale (in space, time or both)
in the space-time by
  means of a DSR-like approach. However, it is clear that --as occurs
  in DSR in the momentum space-- it is always possible to map the
invariant scale (space or time)  to infinity in the classical space
 time and to map the zero of the physical space time to the zero of the
classical  space time. The resulting space-time will differ from the
Lorentz invariant one at large scales but it will not suffer the
problems we discussed above: it will have
a  {\em maximal scale} (and possibly a minimum momentum)
and  will mirror the usual  DSR in the momentum space. 

Finally, let  us say that the statement that
the approach with a minimal scale is not possible, but the one
with a maximal scale is allowed, can be understood by a dimensional
argument. 
If we assume: {\em i}) a continuous differentiable manifold
structure  for the space  time, {\em  ii}) the the existence 
of  a length scale $\ell$, 
it is always possible to express any quantity depending on the
coordinates as a series containing only negative powers of
$\ell$.
If we put the extra condition that {\em iii}) it should exist 
a  smooth limit toward the  undeformed space time, it is clear
the small $\ell$ limit can not be accepted.
 The limit  of large  $\ell$, instead, is  well defined and  the
 interpretation of the  scale as a maximal scale  becomes clear. These
 arguments were already discussed in \cite{referi}.

\section{conclusions}
In this note we have explored a possible scenario for the a space-time
with an invariant scale in a  DSR based approach.
We started constructing a non linear realization of Lorentz
transformations defining a non linear, invertible map between
the physical space and an auxiliary space where the Lorentz group
acts linearly. In doing that we introduce a deformed composition
law for vectors in the physical space to guarantee its invariance
under boosts. 
Up to this point we can still define a translation operator compatible
with the deformed action of boosts, and this 
translations define a group. 

Then we try to impose the physical condition that some (small)
measurable physical length (in space or time) should remain
invariant under boosts. To define the space length we use the standard
expression for the modulus of a vector (again in full analogy with the 
DSR approach in momentum space).
We showed that the invariance requirement is incompatible with 
$i$) a well defined (and invertible) map for all the physical space 
vectors and 
$ii$) with the group structure for the translations since 
the neutral element (and consequently the inverse of any given
translation) can not be unique.

We understand why we encounter differences respect to the 
DSR approach in momentum space. For the latter the invariant 
momentum is realized mapping the physical momentum space up to the 
maximum momentum (energy) to the entire {\em classical} space
($i.e.$ we obtain the invariant scale using the standard Lorentz
fixed point at infinity).
In present case instead, we are forced to map the invariant
scale to the first Lorentz fixed point: the origin
of vector space (recall that both in coordinate and momentum space
the Lorentz transformations are linear and the only two
fixed point are zero and infinity). 
This procedure unavoidably leaves all vectors
with spatial or temporal length smaller or equal to the invariant
length without counterpart in the {\em classical} space. 

The  main  difference of  our  result  with  other approaches  in  the
literature \cite{spacetime},     resides    in    the    definition
of the composition law (\ref{suma}), which has  been introduced  
in order  to  extend the notion  of  covariance. 
We think that the assumption of the standard composition law for vectors 
in the physical space is not correct. 
To assume (\ref{suma})
is an unavoidable step if we want to consistently
construct a non linear realization of Lorentz invariance.

At the end of our construction we arrive to inconsistencies.
A first possibility is, of course, to reject the idea
that a DSR-like transformation can be defined in a coordinate
space-time with an invariant scale. Indeed the connection of
DSR models (defined in momentum space) with coordinate space is unclear
\cite{noi} and may be this connection will be realized only via 
a completely different approach.

In any case, since the non-linear realizations of Lorentz
symmetry have been very successful in constructing new
versions of particle's momentum space, we think it is worthwhile
to explore the same technique further in the attempt to 
understand the possible structure of space-time.

A possible way out is to accept that translations do not form 
any more a group and interpret the peculiar behavior of the
non linear mapping $G$ as a clue for a fundamental indetermination
at the scale of the invariant length.
This interpretation suggests a sort of  (space-time) uncertainty,
something that resembles what is expected to happen in a non
commutative space-time. Alternatively we can speculate that defects (as
in a worm-hole QG vacuum \cite{ruup}) might be present in space-time.


\begin{thebibliography}{99}


\bibitem{foam}Y.Jack Ng,
Lectures given at 40th Winter School of Theoretical Physics: Quantum
Gravity Phenomenology, Ladek Zdroj, Poland, 4-14 Feb 2004, 
{\em gr-qc/0405078}; 
A. Perez, Class. Quant. Grav. {\bf 20}, 43 (2003); 
K. Noui and P. Roche, Class. Quant. Grav.{\bf 20}, 3175 (2003); 
Y. Jack Ng,  Int. J. Mod. Phys. {\bf D11}, 1585 (2002); 
G. Amelino-Camelia, {\em gr-qc/0104005};
J. R. Ellis, N. E. Mavromatos and  D. V. Nanopoulos, 
Phys. Rev. {\bf D63}, 124025 (2001); 
A. Perez and C.  Rovelli, Phys. Rev. {\bf D63}, 041501 (2001); 
L. J. Garay, Int. J. Mod.  Phys. {\bf A14}, 4079 (1999); 
J.  C. Baez, Class. Quant. Grav. {\bf 15}, 1827 (1998).


\bibitem{noncom}
A.  Connes, M.  Douglas and A.  Schwartz, JHEP {\bf 0003}, 9802 (1998); 
N. Seiberg  and E. Witten,  JHEP {\bf 0032}, 9909 (1999);

\bibitem{kamin}
S. Majid and H. Ruegg, Phys. Lett. {\bf B33}, 348 (1993);
S. Zakrzewski, J. Phys. {\bf A 27}, 2075 (1994).


\bibitem{amelino}
G.  Amelino-Camelia, Int. J. Mod. Phys.{\bf D11}, 35 (2002); 
N.R. Bruno, G. Amelino-Camelia and  J. Kowalski-Glikman, 
Phys.   Lett.    {\bf   B522},   133    (2001);   G.   Amelino-Camelia,
Phys. Lett. {\bf B510}, 255 (2001).


\bibitem{smolin}
J. Magueijo and L. Smolin, Phys. Rev Lett. {\bf 88}, 19043 (2002).

\bibitem{mag}
J. Magueijo and L. Smolin, Phys. Rev. {\bf D67}, 044017 (2003).

\bibitem{spacetime}
A.  A.  Deriglazov,  {\em  hep-th/0409232};
S.  Mignemi, {\em  gr-qc/0403038}; 
S.   Gao and  Xiao-ning Wu, {\em  gr-qc/0311009};  
D. Kimberly, J. Magueijo  and J. Medeiros, {\em gr-qc/0303067}. 

\bibitem{KM}
S.  Majid, H.  Ruegg, Phys. Lett. {\bf B334}, 348 (1994);
S. Zakrzewski, J. Phys. {\bf A27}, 2075 (1994). 

\bibitem{KMDSR}
Clemens Heuson,{\em gr-qc/0312034};  
A. Blaut, M. Daszkiewicz, J. Kowalski-Glikman and  S. Nowak, 
Phys. Lett. {\bf B582}, 82 (2004);
J.  Kowalski-Glikman and S.  Nowak, Int. J. Mod. Phys. {\bf D12}, 299
(2003). 

\bibitem{nsim}
G. Amelino-Camelia and M. Arzano, Phys. Rev. {\bf D65}, 084044 (2002); 
G.  Amelino-Camelia, in {\em *Karpacz 2001, New developments in fundamental
interaction theories }, 137 (2001),{\em gr-qc/0106004}. 

\bibitem{rev}
J.   Kowalski-Glikman,  {\em  hep-th/0405273};   G.   Amelino-Camelia,
Int. J. Mod. Phys. {\bf D11}, 1643 (2002).

\bibitem{luk}  
J. Lukierski and A. Nowicki, Int. J. Mod. Phys. {\bf A18}, 7 (2003).

\bibitem{luki}
J.  Lukierski and A.  Nowicki, Int. J. Mod. Phys. {\bf A18}, 7 (2003); 
J.  Lukierski and  A.  Nowicki, Acta Phys. Polon. {\bf B33}, 2537 (2002);
J. Lukierski and A.  Nowicki, Czech. J.  Phys. {\bf 52}, 1261 (2002).


\bibitem{referi} G.   Amelino-Camelia,
Int. J. Mod. Phys. {\bf D11}, 1643 (2002);
G. Amelino-Camelia, {\em 10th Marcel Grossmann Meeting on General Relativity, QG5 session}, Rio de Janeiro, July 20-26, 2003.

\bibitem{noi}
R. Aloisio, A. Galante, A.F. Grillo, E. Luzio, F. Mendez, {\em gr-qc/0410020}.

\bibitem{ruup}
F.  R. Klinkhamer and C. Rupp, Phys. Rev. {\bf D70}, 045020 (2004).

\end{thebibliography}
\end{document}